\documentclass[a4paper, 10pt,twocolumn]{article}

\usepackage[english]{babel}
\usepackage[utf8]{inputenc}
\usepackage[T1]{fontenc}
\usepackage{lmodern}
\usepackage{subfigure}
\usepackage{cite}

\usepackage{graphicx}
\graphicspath{{graphs//}}
\usepackage{amsmath}

\usepackage{setspace}
\usepackage[\ifnum\pdfoutput=1breaklinks\fi]{hyperref}

\title{Observation of photoelectron circular dichroism using a nanosecond laser}
\author{ 
Alexander Kastner, Tom Ring, Hendrike Braun, Arne Senftleben, and Thomas Baumert \\
	Institut für Physik und CINSaT \\
	Universität Kassel, Heinrich-Plett-Strasse 40\\
	34132 Kassel (Germany)  \\
	e-mail: tbaumert@uni-kassel.de \\
	phone: +49 (0)561 804 4452 \ fax: +49 (0)561 804 4453	
	}

\begin{document}
{\centering
\maketitle
}

\textit{ \textbf{Abstract:}
Photoelectron circular dichroism (PECD) is a fascinating phenomenon both from a fundamental science aspect but 
also due to its emerging role as a highly sensitive analytic tool for chiral recognition in the gas phase. PECD has been studied with single-photon as well as multi-photon ionization. The latter has been investigated in the short pulse limit with femtosecond laser pulses, where ionization can be thought of as an instantaneous process. \\ \indent
In this contribution, we demonstrate that multiphoton PECD still can be observed when using an ultra-violet nanosecond pulse to ionize chiral showcase fenchone molecules. Compared to femtosecond ionization, the magnitude of PECD is similar, but the lifetime of intermediate molecular states imprints itself in the photoelectron spectra. Being able to use an industrial nanosecond laser to investigate PECD furthermore reduces the technical requirements to apply PECD in analytical chemistry.
}\\ \vspace{1em}

Chiral recognition in the gas phase with the help of electromagnetic radiation is an emerging research area as an interaction-free environment allows to study intrinsic chiral response. \\ \indent  
In the last few years for instance fully coherent microwave three-wave mixing schemes~\cite{Domingos.2018}, Coulomb explosion imaging for direct absolute configuration determination,~\cite{Pitzer.2018} circular dichroism (CD) in all-optical high-harmonic spectroscopy~\cite{Woerner.2018} as well as CD in ion yield~\cite{Boesl.2016} have been developed. \\ \indent
When chiral molecules are ionized by circularly polarized light, a forward-backward asymmetry in the photelectron emission direction with respect to the light propagation direction was predicted by theory~\cite{Ritchie.1976}. This effect arises in the electric dipole approximation and was first observed experimentally using single-photon ionization~\cite{Bowering.2001}. The forward-backward asymmetry is known as photoelectron circular dichroism~\cite{Powis.2008} (PECD) and typically exhibits large magnitudes with experimental values as high as 37 \%~\cite{Ganjitabar.2018}. The origin of PECD lies in the quantum interference of outgoing partial waves~\cite{Garcia.2013}. As PECD has been observed for single-photon ionization out of an achiral core-shell orbital, it has been interpreted as being mainly a final state effect~\cite{Powis.2008, Lein.2014}. Measuring PECD has been boosted by techniques capturing the full photoelectron momentum distribution  such as velocity map imaging (VMI)~\cite{Chandler.2017} or a recently developed stereo-electron detector~\cite{Miles.2017}. Recently, single-photon PECD~\cite{Nahon.2015, Turchini.2017} highlighting in addition the relevance for astrophysical research~\cite{Hadidi.2018} has been reviewed. \\ \indent
PECD using a table-top femtosecond (fs) laser system has been demonstrated~\cite{Lux.2012, Lehmann.2013} by employing 2+1 resonance-enhanced multi-photon ionization (REMPI). PECD provides the possibility to determine changes in enantiomeric excess (e.e.) down to the sub-one percent regime~\cite{Kastner.2016, Nahon.2016, Comby.2018}, and its analytical capabilities have already been acknowledged~\cite{Boesl.2016, Janssen.2017}. \\ \indent
For REMPI studies, changing the laser wavelength allows to study the dependence of PECD on intermediate state and photoelectron kinetic energy~\cite{Lehmann.2013, Fanood.2016, Beaulieu.2016a, Kastner.2017} in comparison to theoretical descriptions~\cite{Mueller.2018, Goetz.2017}. The influence of polarization state used for excitation and ionization on observed PECD was modeled~\cite{Goetz.2017} and investigated experimentally~\cite{Beaulieu.2018}. Using a fs pump-probe setup, dynamics in the 3s intermediate state were investigated for fenchone~\cite{Beaulieu.2016b, Comby.2016}. When the duration of the laser pulses is increased to longer than picoseconds, intra-molecular dynamics in addition to rotation of the molecules can evolve during ionization. \\ \indent
Here, we report on observation of PECD on the nanosecond (ns) timescale. In this case the lifetime of the intermediate states is much shorter than the laser pulse duration. Ns lasers are widely used in analytical chemistry~\cite{R.E.Russo.2002, Boesl.2016}. The ability to use an ordinary ns laser system to measure PECD furthermore reduces the technical requirements to use PECD in analytics.

\begin{figure}[htb]
\includegraphics[width=\linewidth]{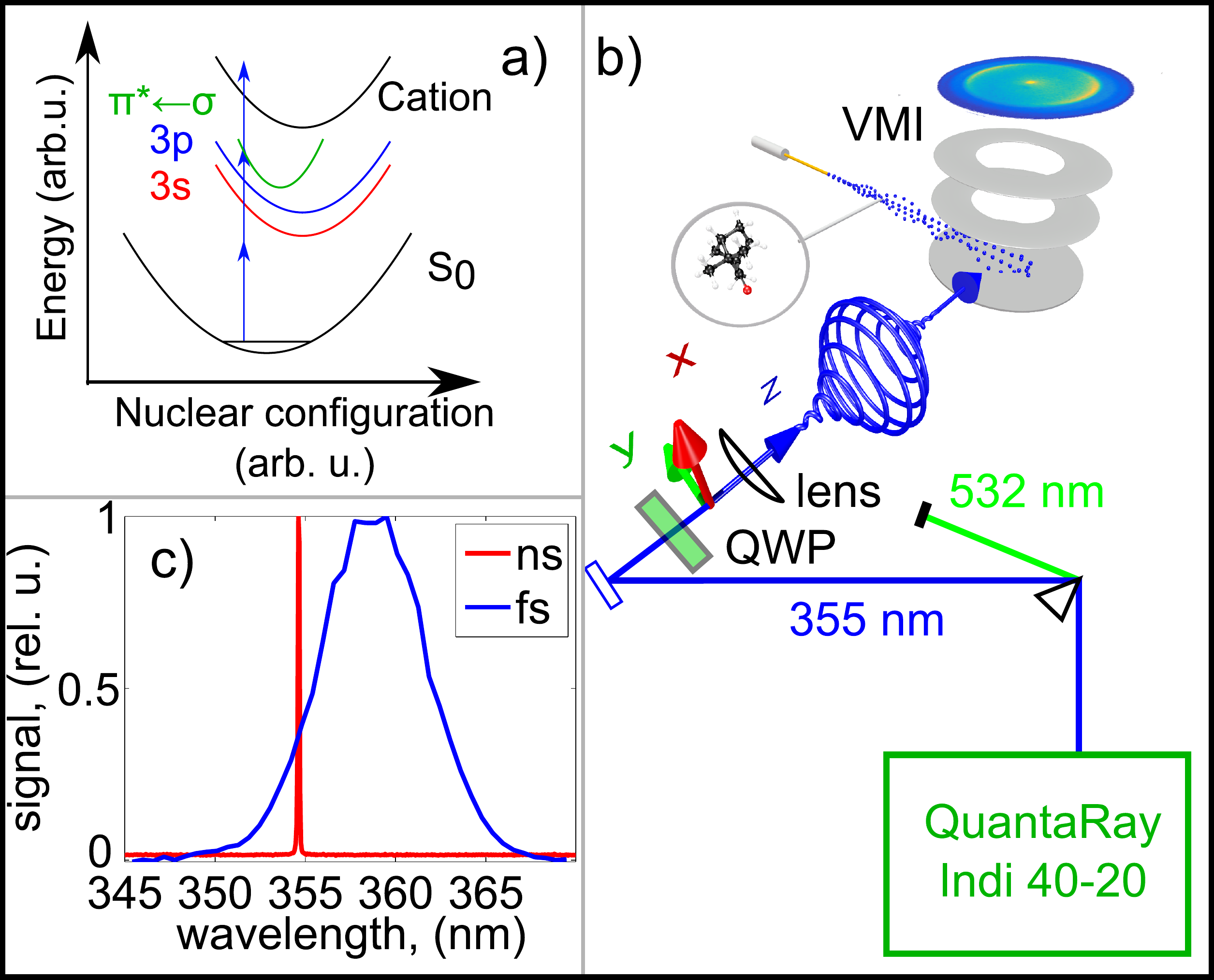}
\caption{\label{fig:Expsetup}
a) 2+1 REMPI scheme observed in previous fs experiments~\cite{Kastner.2017}, b) experimental setup for the ns PECD experiment and c) comparison of ns and fs laser spectra. 
}
\end{figure}

The three-photon excitation and ionization scheme for fenchone derived from a previous fs experiment~\cite{Kastner.2017} is depicted in figure \ref{fig:Expsetup}a). At least three different electronic intermediate states can be populated via two-photon transition from the S$_0$ ground state. The first two correspond to excitation of an electron from the HOMO leading to electronic intermediates having 3s and 3p Rydberg character as observed previously~\cite{Pulm.1997, Driscoll.1991, Kastner.2017}, Assuming parallel potential energy surfaces, the transition from Rydberg state to continuum is governed by a $\Delta v=0$ propensity rule as observed experimentally~\cite{Kastner.2017}. The third transition (due to its different nature not shown in figure \ref{fig:Expsetup}a)) is $\pi^*\leftarrow\sigma$~\cite{Pulm.1997}. \\ \indent
The experimental setup is depicted in figure \ref{fig:Expsetup}b). The results shown herein will be compared to a fs experiment with a spectrum encompassing the ns spectrum (see figure \ref{fig:Expsetup}c)). A detailed description of experimental methods can be found in the Supporting Information (SI). In brief, the ns laser pulses are guided to the VMI chamber and the polarization is converted into either left- or right circularly polarized using a quarter-wave plate (QWP). The laser pulses are focused into the VMI chamber and are intersected with an effusive gas beam of fenchone. The VMI images are analyzed by an expansion into Legendre polynomials, where the odd-order coefficients were weighted and linearly summed to obtain a PECD metric characterized by one number (LPECD)~\cite{Lux.2015}.

\begin{figure}[htb]
\subfigure[\label{fig:PESCompa}]{\includegraphics[width=0.49\linewidth]{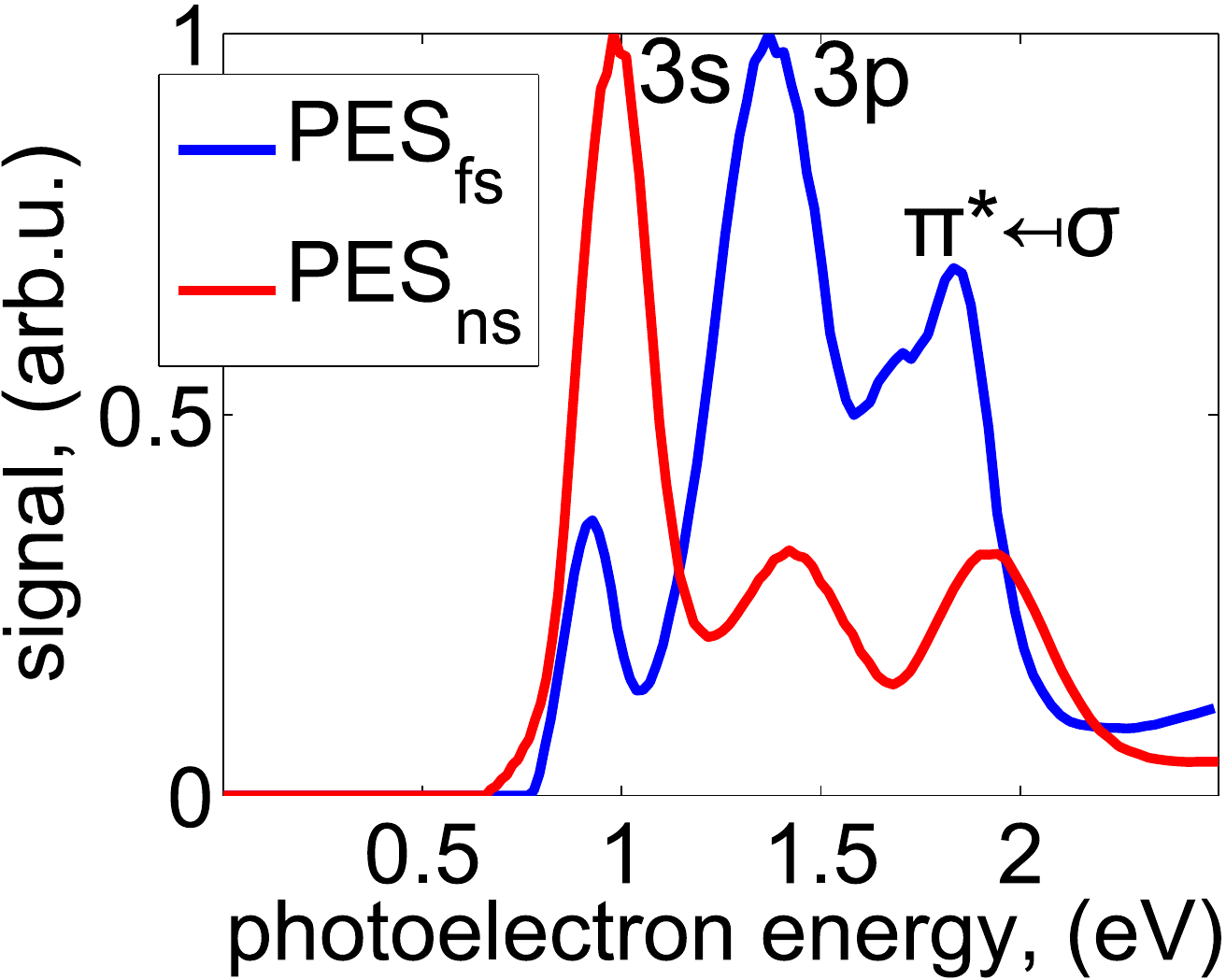}} \hfill
\subfigure[\label{fig:PESCompb}]{\includegraphics[width=0.49\linewidth]{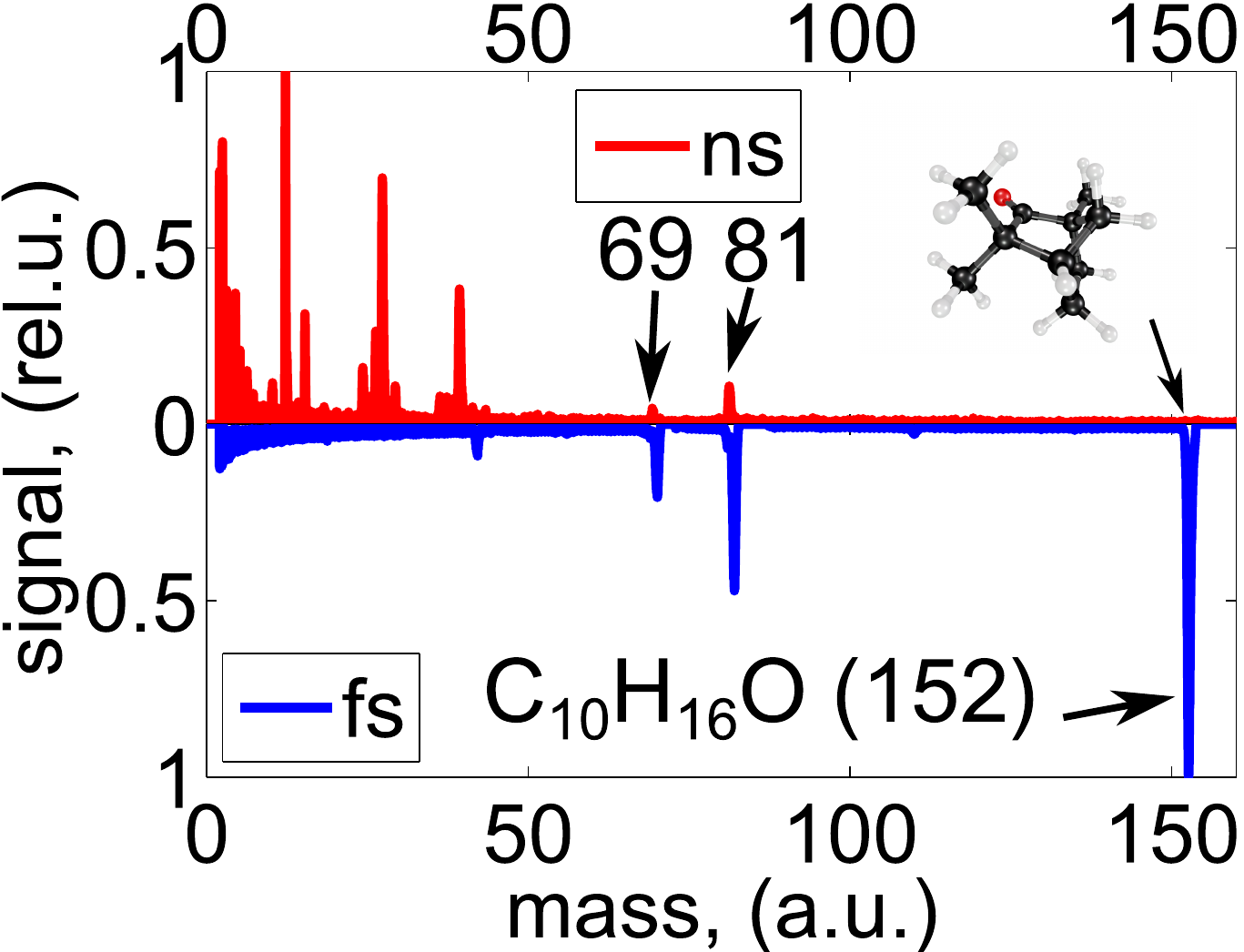}}
\caption{\label{fig:PESComp}
a) Comparison of photoelectron spectra obtained by ionizing (S)-($+$)-fenchone either with the ns laser (red line) or with the fs laser (blue line). b) Ns mass spectrum (red) in comparison with a fs mass spectrum (blue).  
}
\end{figure}

The fs photoelectron spectrum (PES) shown in figure \ref{fig:PESCompa} contains three distinct contributions, which can be assigned to excitation of $\text{3s}\leftarrow\text{n}$, $\text{3p}\leftarrow\text{n}$ and $\pi^*\leftarrow\sigma$ transitions.~\cite{Kastner.2017} The ns PES resembles the fs PES with respect to contributions and their energetic positions. Assuming the formerly observed $\Delta v=0$ propensity rule, the photoelectron energy scales as $\hbar \omega$~\cite{Kastner.2017}. Please note that intra-molecular vibrational energy redistribution as formerly observed for fenchone on a few hundred fs timescale~\cite{Comby.2016, Beaulieu.2016b} may occur in the 3s Rydberg state. Due to the $\Delta v = 0$ propensity rule, the photoelectron energy is not affected by such effects. \\ \indent
Due to $\hbar \omega$ scaling, the Rydberg peaks in the ns PES are up-shifted in energy by the difference in one-photon energy (about $+$40 meV) with respect to the fs PES. Fitting Gaussians underneath the three peaks each for the ns and the fs PES delivers a difference in energy of $+$56 meV for 3s, $+$37 meV for 3p and about \mbox{$+$86 meV} for $\pi^*\leftarrow\sigma$. The first two values are in reasonable agreement with the expected shift due to photon energy. \\ \indent
Earlier experiments reported a picosecond lifetime for the 3s state~\cite{Comby.2016, Kastner.2017} and roughly an order of magnitude shorter lifetime of the 3p state~\cite{Kastner.2017}. In the fs experiment the pulses are short and ionization is fast compared to internal dynamics of the molecule. Therefore, the excitation probability of the intermediate states determines
the respective photoelectron yields. In the ns experiment, the pulse duration exceeds the lifetime of the states. Therefore, the contributions to the PES are influenced in addition by the population decay. Thereby the relative heights in the PES change, as can be noticed in figure \ref{fig:PESCompa}. Most pronounced is the change in relative height of the 3s and 3p contributions. Taking the longer lifetime of the 3s state into account, a longer effective ionization window for the ns laser could lead to a higher signal from this state as compared to the signal of the 3p state. \\ \indent
The mass spectrum for the ns ionization (shown in red in figure \ref{fig:PESCompb}) shows stronger fragmentation of the molecules as compared to the fs experiment (shown in blue). The fs mass spectrum is dominated by the parent ion, whereas in the ns mass spectrum mainly masses below 50 amu are present. The two heaviest fragments with masses 69 amu and 81 amu found in the ns mass spectrum can be found as well in the fs mass spectrum. In previous fs experiments, increasing the laser intensity has been found to lead to more fragmentation while the PECD has stayed basically unaffected~\cite{Lux.2015}. This has been attributed to ionization preceding fragmentation in agreement with coincidence imaging findings~\cite{Lehmann.2013}. In recent mass-tagging experiments on limonene~\cite{Fanood.2016}, barely different PECD values were observed on parent compared to fragment ionization. As ns and fs PES contain the same peaks and the shifts in peak positions can be explained by different center wavelengths, we presume that the ns PES stems mainly from parent ionization prior to ionic dissociation. As we show below, the ns LPECD values are similar compared to fs values. This hints to PECD being recorded from undissociated molecules. \\ \indent

\begin{figure}[htb]
\subfigure[ (S)-($+$)-fenchone \label{fig:PECDAntisSFen}]{\includegraphics[width=0.49\linewidth]{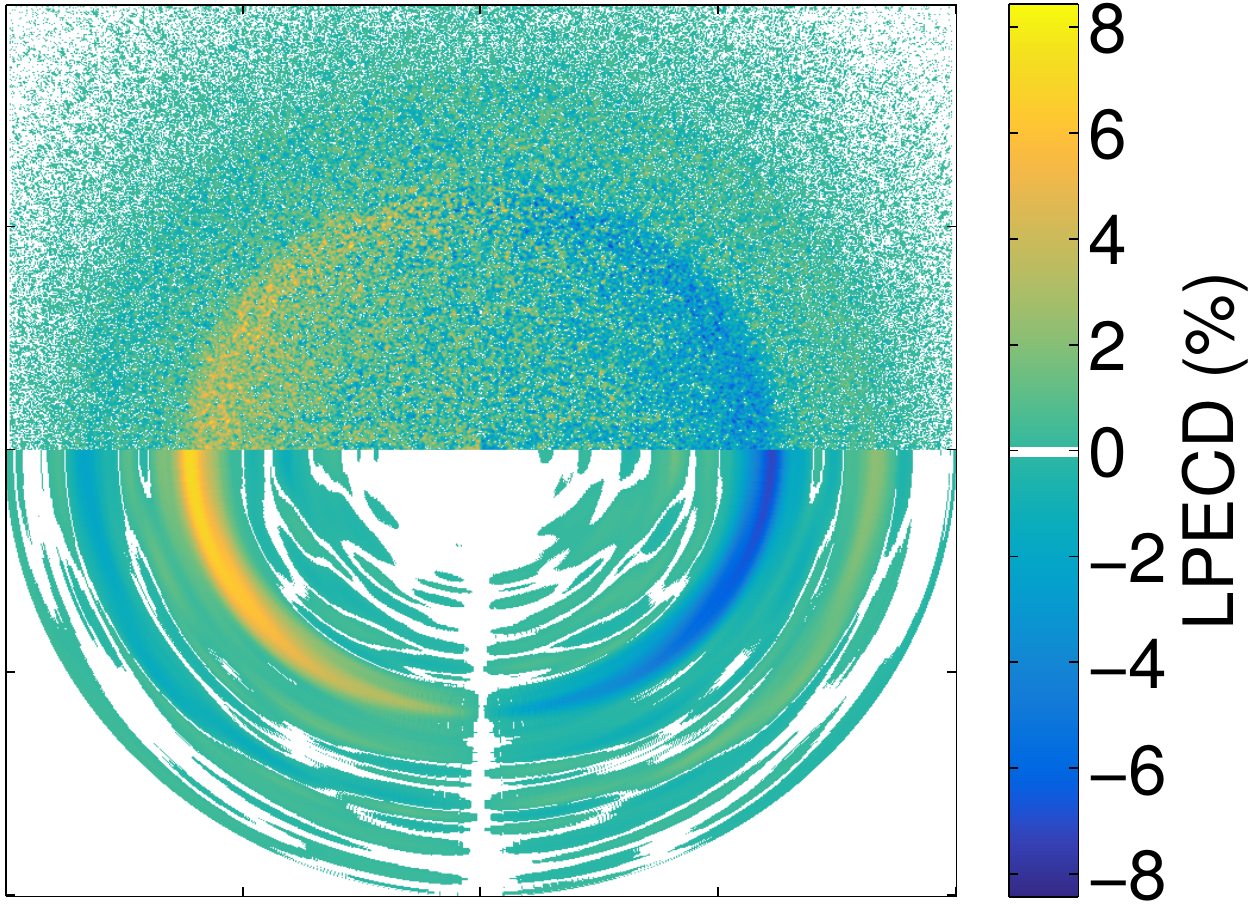}} \hfill
\subfigure[ (R)-($-$)-fenchone \label{fig:PECDAntisRFen}]{\includegraphics[width=0.49\linewidth]{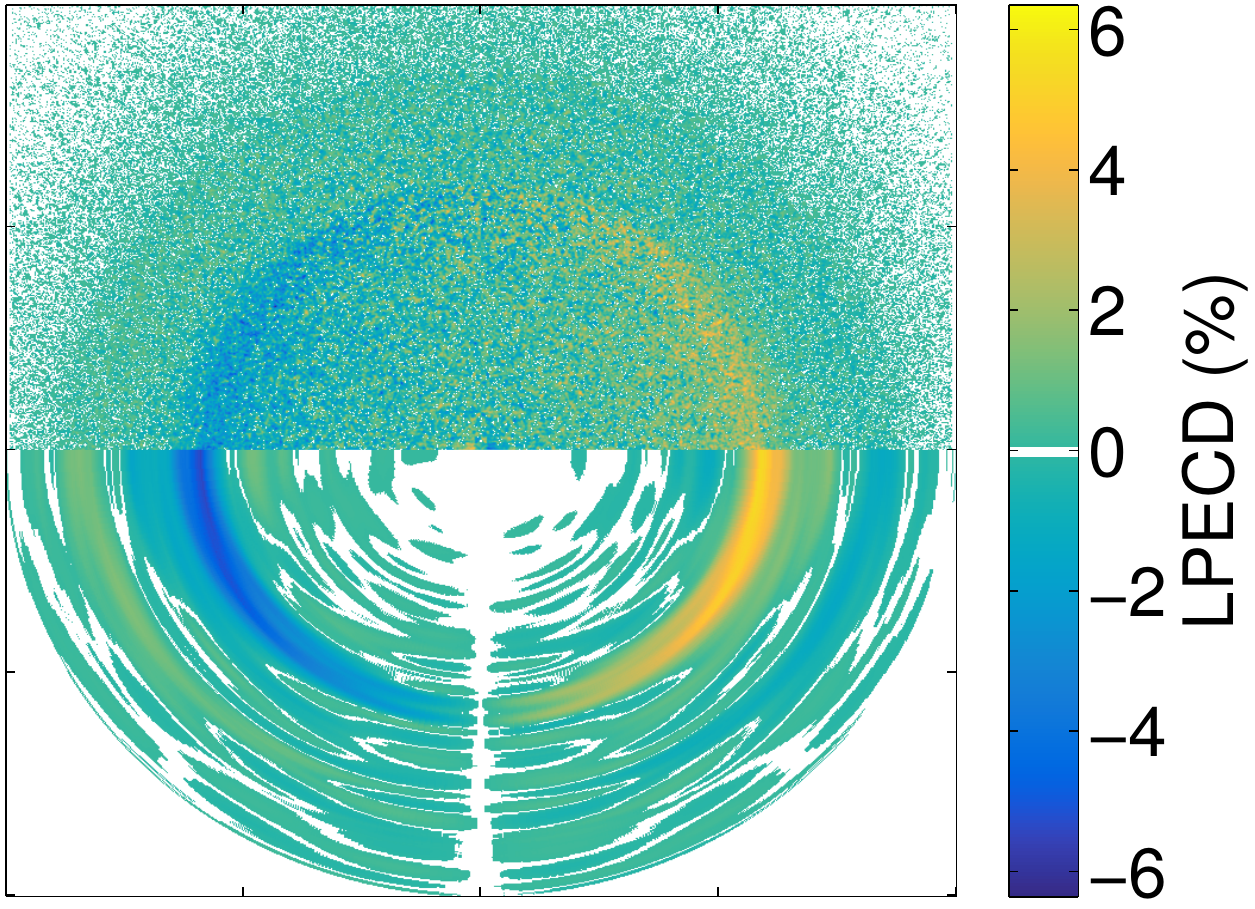}}
\caption{\label{fig:PECDIm}
Anti-symmetrized PECD image for (S)-($+$)-fenchone (a) and (R)-($-$)-fenchone (b). The laser propagates from left to right in the images. 
}
\end{figure}

In figure \ref{fig:PECDIm} the anti-symmetric parts of the PECD images obtained for (S)-($+$)- and (R)-($-$)-fenchone are shown. The raw data is depicted in the upper half of the images, whereas the derived contribution of the odd-order Legendre polynomials is depicted in the lower half of the images. The strongest PECD-bearing contribution is the innermost ring, which comes from excitation of the 3s state. A detailed description of the data evaluation can be found in the SI. In brief, the Legendre coefficients and LPECD values are obtained by weighted averaging over the FWHM of each peak in the PES. In the fs experiment, two contributions to the 3p peak with opposite signs have been observed~\cite{Kastner.2017}. Here, only one contribution underneath the 3p peak is observed. 

\begin{figure}[htb]
\subfigure[\label{fig:LPECD}]{\includegraphics[width=0.49\linewidth]{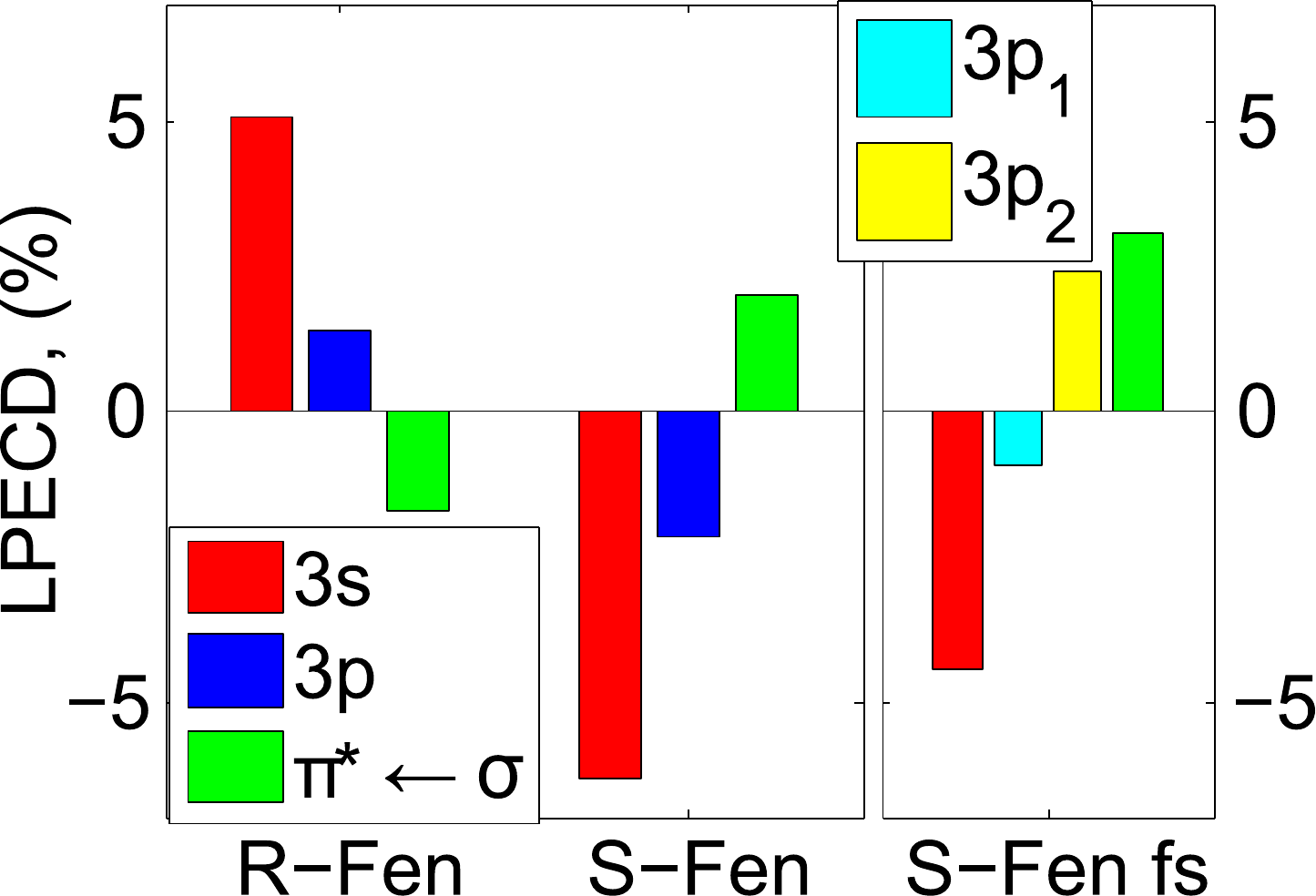}} \hfill
\subfigure[\label{fig:c1}]{\includegraphics[width=0.49\linewidth]{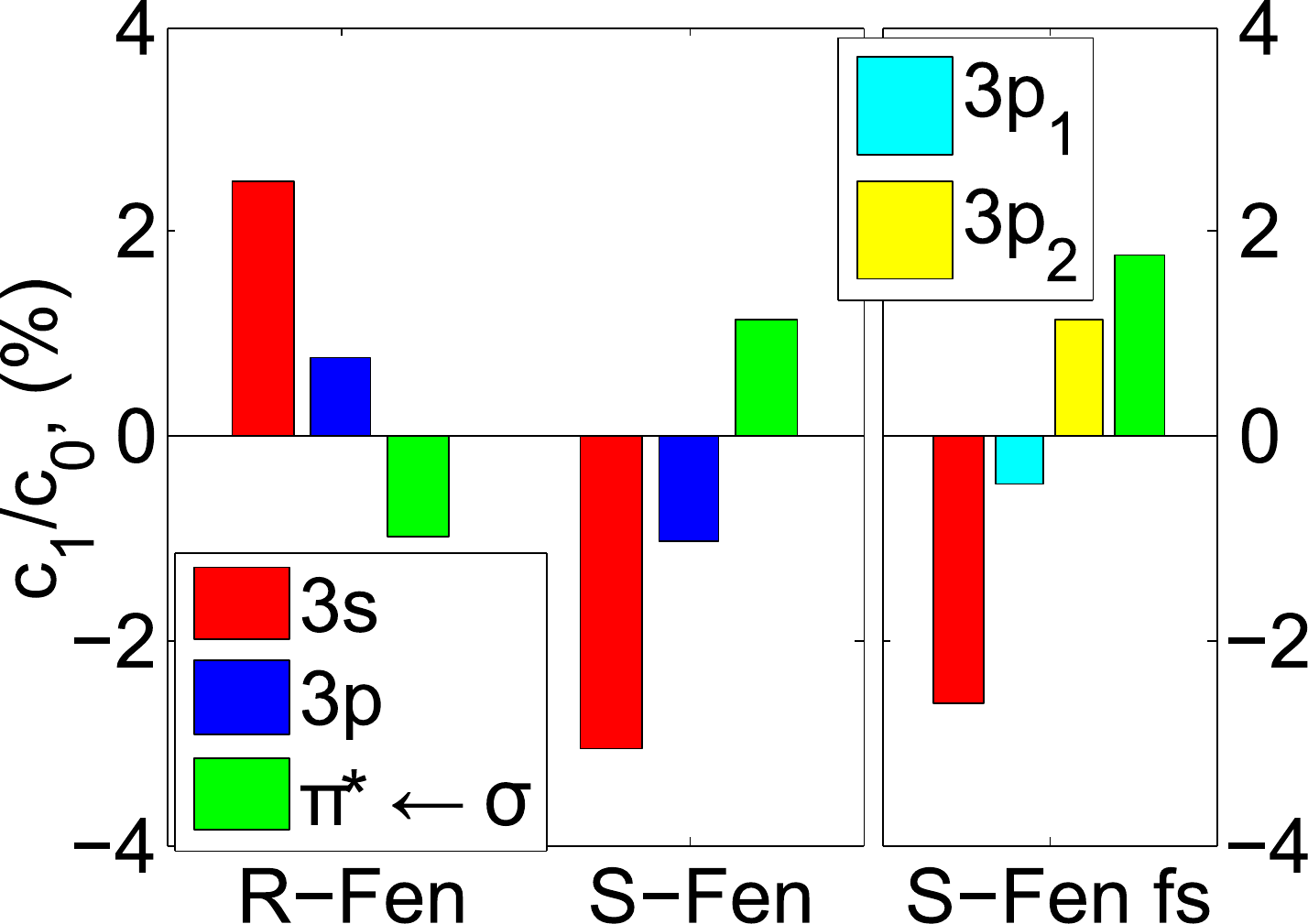}}
\caption{\label{fig:LPECDComp}
Comparison of weighted LPECD (a) and c$_1$/c$_0$ ratio of Legendre coefficients (b) values for the different intermediate resonances (color-coded) for (R)-($-$)-fenchone and (S)-($+$)-fenchone using the ns laser (left two data sets) and for comparison the values obtained for (S)-($+$)-fenchone in a previous fs experiment (right side). 
  }
\end{figure}

The resulting weighted average LPECD values are depicted in figure \ref{fig:LPECD}. Three different data sets each containing the values for the observed contributions to the PES are shown in figure \ref{fig:LPECD}. The two data sets on the left hand side are for both enantiomers of fenchone using the ns laser and the data set shown on the right hand side is data obtained with the fs laser for (S)-($+$)-fenchone. Comparing the ns data, all the LPECD values are mirrored in sign when exchanging the enantiomer, while in magnitude they differ due to the different e.e. for the (R)-($-$)-fenchone specimen (about 84 \% e.e.) compared to the (S)-($+$)-fenchone specimen (about 99.9 \% e.e.)~\cite{Kastner.2016}. Furthermore, for a given enantiomer the sign of LPECD for 3s and 3p is the same, but the sign changes for $\pi^* \leftarrow\sigma$. \\ \indent
If we compare the ns PECD values obtained for (S)-($+$)-fenchone to the ones found in the fs measurement, it can be seen that the magnitude and sign for the 3s as well as $\pi^* \leftarrow\sigma$ contribution is about the same. The value for 3p has the same sign as the 3p$_1$ found in the fs experiment. In the fs experiment, the LPECD for 3p$_2$ has a higher value and opposite sign compared to the 3p$_1$ contribution. This could hint in the direction that several intra-molecular dynamics are launched when exciting the 3p manifold, especially in the energy region labeled 3p$_2$ in the fs experiment~\cite{Kastner.2017}. However, further experimental and theoretical considerations are necessary to clarify this. \\ \indent    
The dominant contributions to the LPECD metric stem from the c$_1$/c$_0$ coefficients in agreement with previous findings~\cite{Kastner.2017}. The c$_1$ values obtained for the different intermediate states for the ns and the fs experiment are plotted in figure \ref{fig:c1} and show reasonable agreement. The higher order coefficients have much weaker influence on the LPECD and are about one order of magnitude smaller compared to the c$_1$ values and are therefore not displayed.  \\ \indent
In this contribution, we demonstrated that PECD can be observed when using an ordinary ns Nd:YAG laser. In comparison to fs findings, the 3s state dominates the PES for the ns experiment attributed to its longer lifetime as compared to the 3p state. The observed LPECD value for the 3s state is in agreement with previous fs PECD values. \\ \indent
Being able to use a commercial ns laser to observe PECD furthermore reduces the technical requirements to apply PECD in analytical chemistry and thus paves the way for many new investigations in chiral recognition in the gas phase.

\section*{Acknowledgement}
The authors thank H. G. Lee for many fruitful discussions. This work was supported by the Deutsche Forschungsgemeinschaft (DFG) within the Sonderforschungsbereich SFB–1319 'Extreme Light for Sensing and Driving Molecular Chirality - ELCH'


\end{document}